\documentclass[aps,prfluids,onecolumn,superscriptaddress,showkeys,longbibliography]{revtex4-2}
\usepackage{bm}
\usepackage{color}
\usepackage{amsmath}  
\usepackage{empheq}
\usepackage{SIunits}
\usepackage[colorlinks=true,allcolors=blue]{hyperref}
\definecolor{vert}{rgb}{0.0, 0.5, 0.0}

\begin{document}

\title{Pilot-wave hydrodynamics of a particle in a density-stratified fluid}

\author{Simon Gsell}
\affiliation{Aix Marseille Univ, CNRS, Centrale Med, IRPHE (UMR 7342), Marseille, France} 

\author{Patrice Le Gal}
\affiliation{Aix Marseille Univ, CNRS, Centrale Med, IRPHE (UMR 7342), Marseille, France} 

\date{\today}

\begin{abstract}

Inspired by bouncing drop experiments that revealed how macroscopic systems can exhibit wave-particle properties previously thought to be exclusive to quantum systems, we introduce here a new wave-particle system based on internal gravity waves propagating in density-stratified fluids. 
Recent experiments on particles (called \textit{ludions}) oscillating in such a fluid medium suggest that wave-particle interactions can induce symmetry breaking, leading to spontaneous self-propulsion of the particle in the horizontal plane.
Here, we propose a minimal hydrodynamic theory showing that this instability can be explained by a Doppler force emerging from interactions between the ludion and its own wave field.
We validate our theoretical predictions using direct numerical simulations, which confirm that the growth of the instability is determined by the particle oscillation amplitude.
In wall-bounded domains, reflections of the internal waves create a Casimir-like potential that rapidly develops and constrains the particle motion.
Despite the presence of the Doppler force, this potential governs the ludion long-term dynamics, leading to capture in fixed points or chaotic attractors near the potential wells.
We show that the essential features of this behavior are well captured by a minimal dynamical model.
Our findings establish the ludion as a novel hydrodynamic pilot-wave system, offering a new platform for exploring macroscopic wave-particle behaviors, particularly in three-dimensional configurations.

\end{abstract}


\maketitle

\section{Introduction}
Wave-particle systems consist of moving particles that interact with the wave field they generate.
Over the past decades, significant attention has been devoted to studying liquid droplets bouncing on a vibrating fluid surface \cite{Couder, CouderandFort,Bush, BushandOza, Bush2}.
In such systems, the droplets interact with the waves they generate, forming a wave-particle system that mimics aspects of quantum wave-particle duality and replicates several quantum-like phenomena on a macroscopic scale \cite{CouderandFort, Pucci2, Anderson, Dubertrand, Eddi, Harris, perrard14}. 
These pioneering works have recently inspired the investigation of other types of macroscopic wave-particle systems, such as self-propelled dipolar acoustic sources \cite{Martischang}, capillary surfers \cite{Rhee, Pucci, Benham2}.
 
Here, we explore the dynamics of a novel wave-particle system consisting of a particle oscillating in a density-stratified fluid. 
Previous experiments \cite{PLG} have shown that such a system can be realized using a Cartesian diver (\textit{a ludion}), which oscillates vertically in response to external pressure oscillations.
These experiments revealed three key phenomena: (i) the emission of internal gravity waves by the oscillating ludion, (ii) spontaneous horizontal motion (swimming) of the ludion, and (iii) complex, unsteady dynamics following the onset of swimming.
However, the underlying physical mechanisms governing the ludion's behavior, particularly the role of the emitted waves in driving its motion, remain unresolved.

\begin{figure} 
  \includegraphics[width=\linewidth]{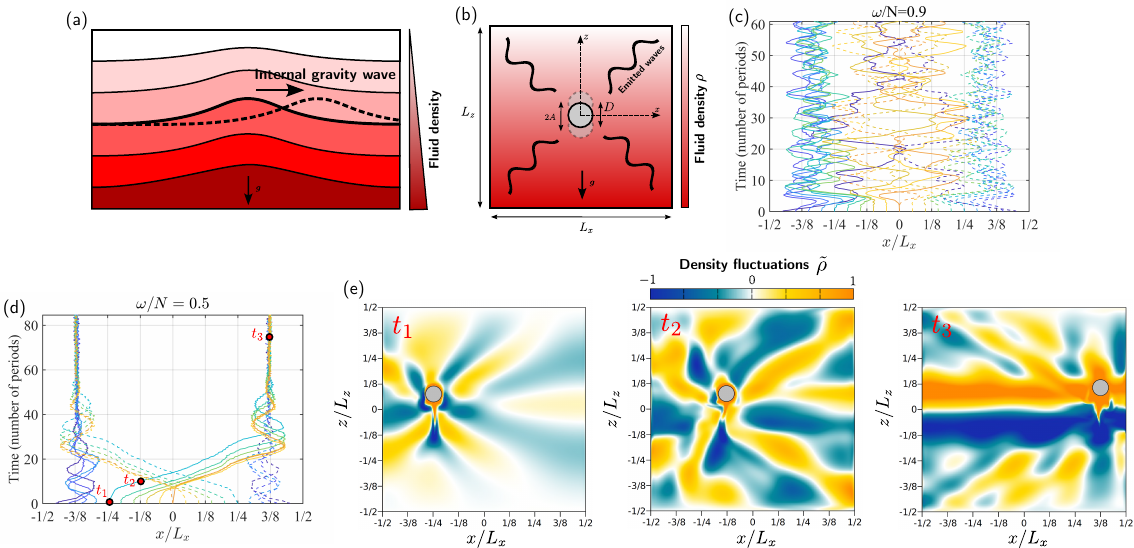}
  \caption{Emergent dynamics of a particle oscillating in a density stratified fluid.
  (a) In a density stratified fluid, perturbations in the vertical density profile propagate as internal gravity waves.
  (b) In a wall-bounded domain, we force a circular particle (the \textit{ludion}) to oscillate vertically in a density stratified fluid, acting as a local source of internal waves.
  (c,d) For a series of initial positions, the particle spontaneously moves in the horizontal direction towards specific attractors, as shown by the trajectories for (c) $\omega/N=0.9$ and (d) $\omega/N=0.5$, with $\omega$ the oscillation pulsation and $N$ the Brunt-Väisälä frequency depending on the density gradient (see also supplementary movies 1 and 2).
  (e) Snapshots showing the development of the normalized fluctuating density field $\tilde{\rho}$ in the fluid domain at three different times (see red dots in (d)) of the ludion dynamics.
  In (c-e), $Re=25$ and $A/D = 1$.
  Only the trajectories starting on the left side of the domain are simulated in practice; trajectories starting on the right side are deduced by symmetry.
  }
  \label{fig:ludion_intro}
\end{figure}

Stably stratified fluids exhibit a vertical density gradient (e.g. due to salinity or temperature variations, as observed in the oceans), with denser fluid at the bottom and lighter fluid at the top (\autoref{fig:ludion_intro}a).
The strength of the stratification is quantified by the Brunt–Väisälä frequency $N$, which represents the inverse of the timescale over which a fluid particle displaced vertically relaxes back to its neutrally buoyant position.
Perturbations to the vertical density profile occurring on timescales longer than $1/N$ lead to the propagation of internal gravity waves \cite{Bruce} (\autoref{fig:ludion_intro}a). 
The oscillating ludion generates such waves \cite{PLG} (\autoref{fig:ludion_intro}b), which can propagate over long distances and interact with surrounding objects. 
While previous studies have examined internal waves generated by moving sources \cite{more23}, including both theoretical \cite{Lighthill,Rarity,Voisin,Bulatov} and experimental approaches \cite{Stevenson,StevensonandThomas}, the coupled dynamics of a freely moving particle interacting with its self-generated internal wave field remains unexplored.

In this work, we combine hydrodynamic theory and numerical simulations to investigate the coupled dynamics of the ludion interacting with its internal wave field. 
We first demonstrate that, consistent with experimental observations, the ludion spontaneously moves in the horizontal direction and exhibits complex dynamics, including chaotic motion and trapping at specific horizontal positions. 
Next, we propose a minimal hydrodynamic theory revealing that any initial velocity difference between the particle and the surrounding fluid is amplified by a Doppler force, providing a mechanism for the observed self-locomotion instability. 
We further characterize the Casimir-like potential \cite{Tarr, Nicolas} that arises when the particle oscillates at varying distances from the walls, creating specific attractors that shape the ludion’s dynamics. 
Finally, we propose a simplified dynamical system that captures the key features of the ludion's behavior. Together, our findings demonstrate how internal gravity waves in stratified fluids can enable the creation of novel three-dimensional hydrodynamic pilot-wave systems.

\section{Emergent dynamics of the ludion}

We simulate the two-dimensional dynamics of a circular particle (\textit{a ludion}) with diameter $D$ and mass $m$, oscillating vertically in a density-stratified fluid (\autoref{fig:ludion_intro} b).
The fluid is contained within a wall-bounded domain of length $L_x$ and height $L_z$.
In contrast with the experimental system \cite{PLG}, here we directly prescribe the vertical particle motion and focus on its emergent dynamics in the horizontal direction.
The vertical position of the particle follows $z_c (t) = A \sin(\omega t)$, where $A$ and $\omega$ denote the amplitude and pulsation of oscillation, respectively.
The particle is free to move in the $x$ direction, governed by the equation of motion
\begin{equation}
  m \dfrac{d^2 x_c}{dt^2} = F_x, 
  \label{eqn:PFD}
\end{equation}
where $F_x$ is the horizontal fluid force including both pressure and shear-stress effects, and $x_c$ is the particle horizontal position.

Initially ($t=0$), the fluid is at rest and vertically stratified with a density profile $\rho_i = \rho_0 + \alpha z$, were $\alpha = \partial \rho_i / \partial z \le 0$ is constant.
The reference density $\rho_0$ denotes the initial fluid density at $z=0$, i.e. the time-averaged vertical position of the ludion. 
We also choose $\rho_0$ as the ludion density, setting the value of $m$ in equation (\ref{eqn:PFD}).
Under Boussinesq approximation, the 2D fluid dynamics is governed by the continuity and Navier-Stokes equations:
\begin{equation}
  \nabla \cdot \textbf{u} = 0
  \label{eqn:divU}
\end{equation}
and
\begin{equation}
  \rho_0 \dfrac{\partial \textbf{u}}{\partial t} + \rho_0 (\textbf{u} \cdot \nabla) \cdot \textbf{u} = - \nabla p + \mu \nabla^2 \textbf{u} + \rho \textbf{g},
  \label{eqn:NS_1}
\end{equation}
where $\mathbf{u} = (u,w)$ is the velocity field, $p$ is the pressure, $\mu$ is the fluid viscosity and $\textbf{g}$ is the gravitational acceleration. 
The density field follows the conservation equation:
\begin{equation}
  \dfrac{\partial \rho}{\partial t} + \nabla \cdot (\rho \textbf{u}) = \beta \nabla^2 \rho, 
  \label{eqn:drho_dt}
\end{equation}
where $\beta$ is the diffusion coefficient.
The system involves 7 parameters: $A, \omega, \rho_0, \beta, \mu, g, D$.
Using Buckingham $\pi$ theorem, the dynamics can be described by 4 non-dimensional parameters: the oscillation Reynolds number $Re_A = \rho_0 \omega A D / 2 \pi \mu$, the frequency ratio $\omega/N$ (where $N= \sqrt{-g \alpha / \rho_0}$ is the Brunt–Väisälä frequency), the Froude number $Fr = \omega A / N D$ and the Schmidt number $Sc = \mu / \rho_0 \beta$.
In all simulations, we fix $Sc=100$, $D=40 \Delta x$ (with $\Delta x$ the mesh spacing), and a spatial resolution of $512 \times 512$. 
Details of the numerical methods are provided in Appendix A.
We simulate the ludion dynamics in a wall-bounded domain for $A/D=1$ and various initial positions and frequency ratios (\autoref{fig:ludion_intro}, supplementary movies 1 and 2). 
For all the explored frequencies $\omega / N \in ]0,1[$, the ludion spontaneously moves along the $x$ direction after a few oscillation cycles (\autoref{fig:ludion_intro} c-d).
As it oscillates vertically, the ludion emits gravity waves exhibiting their typical oblique beam pattern \cite{Bruce} (\autoref{fig:ludion_intro} e-left).
Over long time scales, the waves strongly interact with the walls (\autoref{fig:ludion_intro} e-right).
However, even when the particle reaches a stable position, the resulting wave pattern does not form a simple chessboard eigenmode pattern, but instead exhibits a complex mix of oblique beams, horizontal layering and stationary waves.

For $\omega / N = 0.5$, the ludion robustly converges to two stable positions located at $x_s= \pm 3 L_x / 8$, regardless of its initial position (see \autoref{fig:ludion_intro} d). 
Attractors are also observed for $\omega / N = 0.9$, but in this case, the ludion does not reach stable positions and instead exhibits chaotic meandering near $x=0$ or $x_s$ (see \autoref{fig:ludion_intro} c).
These examples illustrate the typical dynamics observed across a range of forcing frequencies ($\omega /N$) and amplitudes ($A/D$), characterized by three key features: (i) spontaneous horizontal motions, (ii) attraction to stable positions or (iii) chaotic motions. 
Across all cases, the ludion swimming velocity is consistently of the order of $U_\infty=\omega D/ 20$.

\section{Dispersion relation}
\begin{figure}[!t]
  \includegraphics[width=\linewidth]{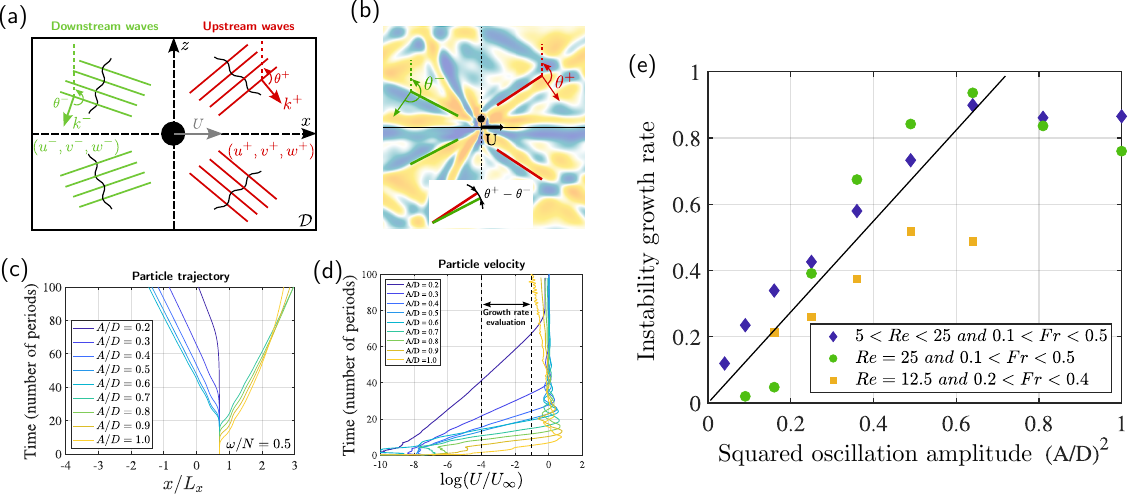}
  \vskip-0.5em
  \caption{Symmetry breaking of the wave pattern and emergence of a Doppler force acting on the ludion.
  (a) An oscillating moving source emits a pair of planar waves with wave vector $k^+$ (upstream) and $k^-$ (downstream), respectively forming angles $\theta^+$ and $\theta^-$ with the vertical axis.
  (b) Numerical illustration of the asymetric wave pattern in the vicinity of the moving ludion.
  (c) Trajectories of a ludion in a horizontally-periodic domain, for $\omega/N = 0.5$ and various oscillation amplitudes (see also supplementary movies 3 and 4).
  (d) Log-lin plot of the time evolution of the normalized ludion velocity, with $U_\infty=\omega D/ 20$, for trajectories shown in (c), confirming the exponential growth of the ludion velocity until it saturates to the constant velocity $U_\infty$.
  (e) Growth rate of the ludion velocity as a function of $(A/D)^2$ for different values of Reynolds and Froude numbers, reproducing the linear trend predicted by our theoretical model, equation (\ref{eqn:momentumint4}).}
  \label{fig:doppler}
\end{figure}

From the linearized 2D Euler equation (\ref{eqn:NS_1}), the mass conservation equation (\ref{eqn:drho_dt}) and using the Boussinesq hypothesis, one can show that the fluid dynamics is governed by a single partial derivative equation \cite{Bruce}.
Here, we perform the same calculation in the reference frame of a particle moving at a constant velocity $U>0$ in the $x$ direction.
In this case, the equation of motion reduces to the following single equation for the vertical velocity component $w$:
\begin{equation}
  \left( \frac{\partial}{\partial t} - U \frac{\partial}{\partial x} \right) ^2 \left( \frac{\partial ^2}{\partial x ^2} + \frac{\partial ^2}{\partial z ^2} \right) w + N^2 \frac{\partial ^2}{\partial x ^2} w = 0.
 \label{eqn:Euler}
  \end{equation}
Assuming that the swimming speed $U$ is small compared to the typical wave velocity, we consider plane wave solutions of the form $w = \hat{w} e^{i(kx+qz-\omega t)}$, where $k$ and $q$ are the wave vectors in the $x$ and $z$ directions, and $\hat{w}$ is the wave amplitude.
Substituting into equation (\ref{eqn:Euler}) leads to the dispersion relation governing internal waves emitted by the moving source \cite{Lighthill,Rarity}:
\begin{equation}
(\omega + k U)^2 (k^2 +q^2) = N^2 k^2.
\label{eqn:Fourier}
\end{equation}
This equation defines two distinct wave solutions propagating upstream and downstream relative to the moving source, depending on the sign of $k$ (\autoref{fig:doppler} a). 
We denote the upstream-traveling waves by the wave vector $\mathbf{k^+}= ({k}^{+},{q}^{+})$ at an angle $\theta^+$ with respect to the vertical. 
Conversely, $\mathbf{k^-}=({k}^{-},{q}^{-})$ and $\theta^-$ characterize the downstream-propagating waves.
Introducing the wavevector norm $\kappa$ such that ${k^+ = \kappa \sin(\theta^+)}$ and ${k^- = \kappa \sin(\theta^-)}$, the dispersion relation (\ref{eqn:Fourier}) yields
\begin{equation}
{\sin(\theta^+)} = \frac{-\omega}{\kappa U-N},~~\textnormal{and}~~  {\sin(\theta^-)} = \frac{-\omega}{\kappa U+N}.
\label{eqn:Angle2}
\end{equation}
Assuming an invariant wavelength $2\pi/\kappa$ in both upstream and downstream directions, the principal effect of the horizontal motion of the source at speed $U$ is to rotate the wave vectors in the $(x,z)$ plane.
Specifically, as $U$ increases, $\sin(\theta^+)$ increases, causing $\theta^+$ to decrease (since $\theta^+ > \pi/2$). 
Conversely, $\theta^-$ decreases with increasing $U$ (see \autoref{fig:doppler}a,b). 
As shown in the following, this symmetry breaking leads to a net force that amplifies the particle velocity.

\section{Non-linear wave analysis}
The dispersion relation (\ref{eqn:Fourier}) describes the propagation of planar waves far from the moving particle. 
However, in the vicinity of the particle, the wave solution is generally nonlinear and non-planar.
Following a classical non-linear analysis approach \cite{Bruce,Ansong}, we expand the velocity and pressure fields as a power series in a small parameter $\epsilon$ (typically, here, the particle horizontal velocity), $(u,w,p)= \epsilon (u_1,w_1,p_1) + \epsilon^2 (u_2,w_2,p_2) +...$, where $(u,w,p)$ represent the horizontal velocity, vertical velocity and pressure fields, respectively.
The first-order terms $(u_1,w_1,p_1)$ correspond to the linear wave solution, while $(u_2,w_2,p_2)$ represent the unknown second-order contribution.
Substituting this expansion into the Euler equations (\ref{eqn:Euler}) and using the incompressibility condition (\ref{eqn:divU}) leads, at second order, to the horizontal momentum equation,
\begin{equation} 
\rho_0 \frac{\partial u_2}{\partial t} + \rho_0 \left( \frac{\partial (u_1 u_1)}{\partial x} + \frac{\partial (u_1 w_1)}{\partial z} \right) = - \frac{\partial p_2}{\partial x} + f_{L \rightarrow F},
\label{eqn:momentum} 
\end{equation}
where $ f_{L \rightarrow F}$ is a local momentum source accounting for fluid-particle interactions and representing the force exerted by the particle on the fluid.
The time-averaged momentum balance over the entire fluid volume $\mathcal{D}$ is then 
\begin{equation} 
  \begin{split}
  &\rho_0 \iint_{\mathcal{D}}\left( \langle\frac{\partial (u_1 u_1)}{\partial x}\rangle + \langle\frac{\partial (u_1 w_1)}{\partial z}\rangle \right)~dxdz \\
  &= - \iint_{\mathcal{D}} \langle\frac{\partial p_2}{\partial x}\rangle ~dxdz + F_{L \rightarrow F},
  \end{split}
  \label{eqn:momentumint} 
  \end{equation}
where $\langle.\rangle$ denotes the time-averaging operator and $F_{L \rightarrow F}$ is the time-averaged force exerted by the ludion. 
The time derivative in equation (\ref{eqn:momentum}) vanished in equation [\ref{eqn:momentumint}] due to time averaging.
Using the divergence theorem, volume integrals in equation (\ref{eqn:momentumint}) are replaced by integrals over the domain boundaries, namely 
\begin{equation} 
  \begin{split}
  &\rho_0 \int_{-H/2}^{H/2}\left(\langle u_1^{+} u_1^{+}\rangle - \langle u_1^{-} u_1^{-}\rangle \right) dz \\
  =&  \int_{-H/2}^{H/2}\left(\langle p_2^+\rangle - \langle p_2^-\rangle\right) dz +F_{L \rightarrow F},
  \end{split}
  \label{eqn:momentumint2} 
\end{equation}
where superscripts $+$ and $-$ refer to the upstream and downstream vertical boundaries of the rectangular fluid domain with height $H$, and integrals over horizontal boundaries vanished due to top-down symmetry. 
As stated above, we assume the wave solution far from the ludion to recover the linear/planar wave solution.
In a large enough domain $\mathcal{D}$, $p_2$ contributions in equation (\ref{eqn:momentumint2}) thus vanish, and the momentum balance equation reduces to
\begin{equation} 
  \rho_0 \int_{-H/2}^{H/2}\left( \langle u_1^{+} u_1^{+}\rangle - \langle u_1^{-} u_1^{-}\rangle \right) dz=  F_{L \rightarrow F}.
  \label{eqn:momentumint3} 
\end{equation}
This equation states that any asymmetry in the momentum fluxes $\langle u_1^{+} u_1^{+}\rangle$ and $\langle u_1^{-} u_1^{-}\rangle$ must be counterbalanced by a force exerted by the particle on the fluid.

\section{Doppler force and swimming instability}

The upstream and downstream momentum fluxes in equation (\ref{eqn:momentumint3}) can be computed using the first-order wave solution \cite{Bruce} ($u_1 = \hat{u} e^{i(kx+qz-\omega t)} = - \frac{q}{k} \hat{w} e^{i(kx+qz-\omega t)}$, with $(k,q) = (k^+,q^+)$ or $(k^-, q^-)$), yielding
\begin{equation} 
  F_{L \rightarrow F} = \frac{\rho_0 \hat{w}^2 H}{2} \left( \frac{1}{\sin^2(\theta^+)}-\frac{1}{\sin^2(\theta^-)} \right).
\label{eqn:flux} 
\end{equation}
Substituting the expressions of $\sin(\theta^+)$ and $\sin(\theta^-)$ from the dispersion relation (\ref{eqn:Angle2}), 
the force exerted by the fluid on the particle simplifies to
\begin{equation}
  F_{F \rightarrow L}  = \frac{2 \rho_0 H \hat{w}^2 N^2 \kappa N}{\omega^2}U = (2 \rho \kappa A^2 H N) U,
   \label{eqn:momentumint4} 
 \end{equation}
 with $\hat{w} = A \omega$.
This force destabilizes an initially stationary particle, causing its velocity to grow exponentially over time, with a growth rate proportional to the square of the oscillation amplitude, $A^2$. 
Eventually, we expect viscous effects to balance the Doppler force, leading to a steady terminal velocity.

Numerical simulations in horizontally periodic domains confirm this phenomenon: the particle spontaneously moves in the $x$-direction (\autoref{fig:doppler}c), with an initial exponential velocity growth before saturating at a characteristic velocity $U_{\infty}$ (\autoref{fig:doppler}d). 
The measured growth rates scale quadratically with $A$ (\autoref{fig:doppler}e), in agreement with theory. Furthermore, additional simulations have confirmed that when $\omega / N > 1$—i.e., when no internal waves are emitted—the particle remains stationary.

\begin{figure*} 
  \includegraphics[width=\linewidth]{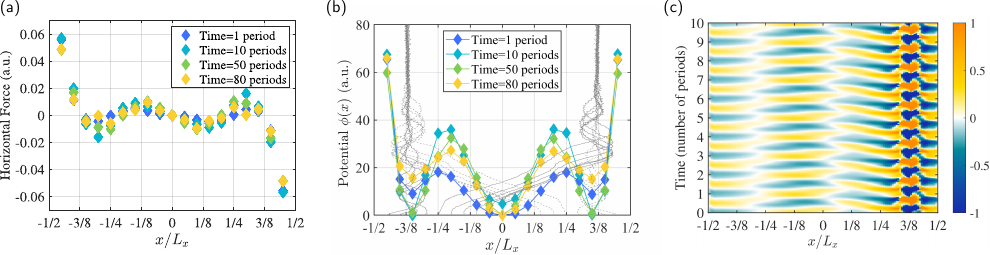}
  \vskip-0.5em
  \caption{Casimir-like potential in a wall-bounded domain.
  (a) We perform a quasi-static analysis by measuring the time-averaged hydrodynamic force applied on a ludion fixed at different horizontal positions in a wall-bounded domain. 
  The force is averaged over $4$ oscillation periods, and we plot the force value at different times after the onset of oscillations.
  (b) Casimir like potential $\phi(x)$ determined by integrating the force distribution in (a). 
  We superimpose the ludion trajectories observed when the ludion is free to swim (\autoref{fig:ludion_intro} b), showing that attractors match with the two side wells of $\phi(x)$. 
  In all panels, $\omega/N = 0.5$ and $A/D=1$. 
  (c) Space-time diagram of the dimensionless vertical fluid velocity $w / \omega D$ along the $z=0$ horizontal line when the ludion is trapped in the right-side well.
  }
  \label{fig:casimir}
\end{figure*}

\section{Casimir-like potential in wall-bounded domains}

The internal waves emitted by the ludion can interact with distant objects and significantly influence its dynamics.
To characterize this effect, we now investigate motion-independent forces acting on a horizontally static ludion in a closed domain.
This is done through a quasi-static analysis, where we measure the time-averaged hydrodynamic force exerted on the particle for different fixed positions along the horizontal axis (\autoref{fig:casimir}a).

Even though the wave field considered here is generated by the particle itself, the emergence of this force is reminiscent of the Casimir effect, in which a radiation pressure—whether electromagnetic \cite{Casimir48}, acoustic \cite{Casimiracoustic}, or hydrodynamic \cite{Casimirmaritime, Tarr, Nicolas}—induces forces on objects or walls embedded in the wave system.
To quantify this phenomenon, we integrate the measured force along the $x$ direction to compute a Casimir-like potential $\phi(x)$, shown in \autoref{fig:casimir}b. 
While the potential $\phi(x)$ exhibits small temporal fluctuations, it rapidly stabilizes and consistently develops three wells: one at the center ($x/L_x = 0$) and two near the vertical walls at approximately $x_s = \pm 3 L_x / 8$.
Notably, when the ludion is free to move, its stable equilibrium positions align precisely with the minima of these side wells, as evidenced by the convergence of trajectories in \autoref{fig:ludion_intro}b.
Interestingly, ludions initially placed near the center of the domain can escape the central well due to the combined effects of the Casimir and Doppler forces.
As shown in \autoref{fig:casimir}b, the ludion also decelerates while passing through the saddle points of $\phi(x)$, around $x/L_x \simeq 1/4$.
Once trapped in a potential well, a standing wave appears along the $z=0$ horizontal line (\autoref{fig:casimir}c).
However, directly linking the potential wells to specific wave patterns remains challenging, as the wave field is inherently complex and two-dimensional (\autoref{fig:ludion_intro}c).

While previous experimental results on ludion dynamics have been interpreted through a bifurcation diagram \cite{PLG}, here our analysis of the Casimir-like potential in a wall-bounded domain suggests that the particle only has a finite number of equilibrium positions in the domain. In the present example for $\omega / N = 0.5$, the only unstable equilibrium position is at the center of the domain ($x=0$). However, over long time scales, the ludion eventually gets trapped in the stable potential wells, hence there is no nonzero, finite terminal velocity that would allow the construction of a bifurcation diagram.

\begin{figure}
\centering 
 \includegraphics[width=0.9 \linewidth]{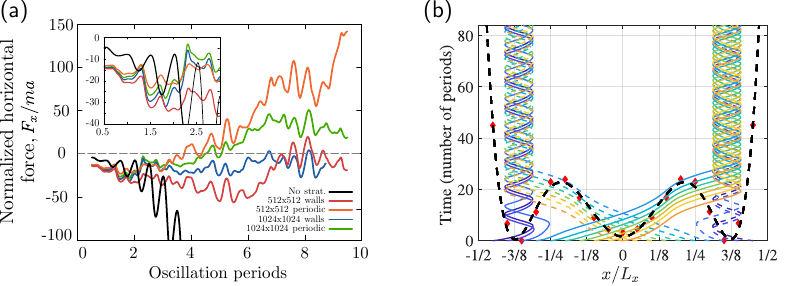}
  \vskip-0.5em
    \caption{
    (a) We characterize the fluid-induced added mass of the ludion by simulating an impulse start at a constant acceleration $a = U_\infty / T_y$, with $T_y$ the oscillation period in the $y$ direction.
    The time history of the horizontal force (averaged over one oscillation period) is plotted for different boundary conditions, normalized by $ma$.
    A case without stratification is also included for comparison.
    (b) Ludion dynamics predicted by the minimal dynamical model (\ref{eqn:Systdyn}), with $C_a=25$, for $A/D=1$ and $\omega / N = 0.5$. The red symbols and the black dash line are the measured potential values at time 10 T and their sixth order polynomial fit.
    }
  \label{fig:TrajSystDyn}
\end{figure}

\section{Minimal dynamical model}

Our findings indicate that the ludion dynamics arise from the interplay of three distinct mechanisms: (i) a Doppler force, which drives the ludion out of equilibrium; (ii) a viscous drag, necessary for the ludion to reach a terminal velocity in an open domain; and (iii) a Casimir-like potential, which attracts the ludion toward geometry-dependent equilibria.
These effects allow us to formulate a minimal dynamical model governing the ludion motion along the $x$-direction:
\begin{equation}
  m (1+C_{a}) \frac{d^2x_c}{dt^2}= \delta \frac{dx_c}{dt} - \gamma \left|{\frac{dx_c}{dt}}\right| \frac{dx_c}{dt} -\frac{d\phi}{dx} ,      \label{eqn:Systdyn} 
\end{equation}
where $m_e = m (1+C_{a})$ is the effective mass of the ludion, $C_{a}$ is an unknown added mass coefficient and $(\delta, \gamma)$ are swimming and drag coefficients to be determined.
The three terms on the right hand side of equation (\ref{eqn:Systdyn}) correspond to the mechanisms described above. 
We model the viscous drag as quadratic in velocity, leading to a terminal velocity in an open domain given by $U_\infty = \delta / \gamma$, which is set equal to the terminal velocity observed in simulations, $U_\infty = \omega D/20$.
The coefficient $\delta$ is related to the amplitude-dependent growth rate characterized in \autoref{fig:doppler}e.
From the linear fit, we estimate $\delta / m_e \approx 1.5 (A/D)^2$.
To determine the last unknown parameter, $m_e$, we perform additional simulations in which the oscillating ludion, initially at rest in the $x$ direction, is subjected to an impulse start with constant horizontal acceleration comparable to that observed in \autoref{fig:doppler}.
For different domain sizes and boundary conditions, we find that the fluid-induced inertial force exerted on the ludion when it starts to move is significantly amplified in the presence of stratification (\autoref{fig:TrajSystDyn}a).
The measured initial force value suggests an added mass coefficient in the range $C_a \in [10,30]$.
By incorporating these parameters and selecting $C_a = 25$, the minimal dynamical model (\ref{eqn:Systdyn}) produces trajectories in reasonable agreement with the simulated dynamics for $A/D = 1$ and $\omega / N = 0.5$ (\autoref{fig:TrajSystDyn}b and \autoref{fig:ludion_intro}b).
However, the model also predicts steady oscillations around the two potential wells, that are not observed in the simulations.
A closer look at equation (\ref{eqn:Systdyn}) indeed shows that a particle, initially placed at the center of a locally quadratic potential well $\phi \sim x^2$, is unstable, hence the observed oscillations.
The absence of such oscillations in our direct simulations suggests the existence of more complex damping effects in the full Navier-Stokes system.

\section{Discussion}
\label{sec:discussio}

Since the pioneering experiments of Couder and Fort on bouncing droplets \cite{Couder}, the search for possible macroscopic pilot-wave systems has become a major scientific challenge.
Here, we introduce a new system driven by internal gravity waves propagating in density-stratified fluids.
We demonstrate that an oscillating particle, called a \textit{ludion}, undergoes horizontal motion due to a wave-induced Doppler force, which amplifies any initial velocity.
Additionally, the internal waves emitted by the ludion interact with distant boundaries, giving rise to a Casimir-like potential that defines specific attractors for the ludion dynamics.

This system can be experimentally realized using a Cartesian diver oscillating vertically in response to external pressure variations \cite{PLG}.
Such experimental system however differs from the present numerical model in the way vertical motion is controlled, as a two-way coupling between horizontal and vertical motion may affect the resulting dynamics, e.g. through complex interactions between buoyancy, added mass and drag \cite{Magnaudet,more23}.
Yet we expect the main conclusions of this work to apply qualitatively to experimental systems. 
First, the Doppler force mechanism is highly general. 
According to our theoretical model (\ref{eqn:momentumint4}), possible modulations of the vertical amplitude are expected to affect the force amplitude, but not the instability leading to self-locomotion.
Similarly, the existence of a Casimir-like potential and of specific equilibrium positions within a wall-bounded domain is not expected to be qualitatively affected by the ludion vertical motion.

The ludion system offers several advantages over existing pilot-wave systems.
First, the Doppler force driving the ludion motion emerges over a broad range of forcing frequencies ($\omega \in ]0, N[$, where $N$ is the Brunt–Väisälä frequency characterizing the density stratification), rather than being restricted to a specific resonance region.
Second, the quadratic dependence of the Doppler force amplitude on the oscillation amplitude enables real-time control of ludion locomotion.
Third, the Casimir-like potential develops rapidly compared to the timescale of the ludion oscillations and interacts with distant objects, potentially leading to pseudo-nonlocal effects.
Finally, unlike bouncing droplet systems that interact with two-dimensional surface waves, the ludion intrinsically couples to a three-dimensional wave field.

Several open questions and extensions arise from this study.
A fundamental understanding of how wave dynamics generate the Casimir-like potential remains lacking.
In particular, the wave patterns associated with potential wells do not match the simple sinusoidal eigenmodes of a square domain, as one might expect by analogy with bouncing droplet systems \cite{Harris}.
Additionally, our minimal dynamical model assumes a constant effective mass for the ludion and a simple quadratic dissipation force.
More refined models may be required to capture complex features such as chaotic motion and the damping of oscillations near potential wells.
In the longer term, we believe this work opens new avenues for studying macroscopic pilot-wave phenomena, including three-dimensional effects, interactions with external potentials or obstacles, and the collective dynamics of ludion swarms.

\section*{Acknowledgments}
The Centre de Calcul Intensif d’Aix-Marseille is acknowledged for granting access to its high performance computing resources.

\section*{Appendix A: Computational model}
\label{appendix_comp}

Introducing the density variation $\tilde{\rho} = \rho - \rho_i$, equation (\ref{eqn:NS_1}) can be written as
\begin{equation}
  \rho_0 \dfrac{\partial\textbf{u}}{\partial t} + 
 \rho_0 (\textbf{u} \cdot \nabla) \cdot {\textbf{u}} = - \nabla P^* + \mu \nabla^2 {\textbf{u}} - \tilde{\rho} {\textbf{g}},
  \label{eqn:NS_2}
\end{equation}
with the new pressure field $P^* = P + P'$ and $P' = \rho_0 z + \alpha z^2 / 2$ a hydrostatic pressure field resulting from the initial density stratification.
Using the density variation $\tilde{\rho}$, the density dynamics equation becomes
\begin{equation}
 \dfrac{\partial \tilde{\rho}}{\partial t} + \nabla \cdot (\tilde{\rho} \textbf{u}) = \beta\nabla^2 \tilde{\rho} - \alpha w. 
 \label{eqn:drho_dt_2}
\end{equation}

We simulate equations (\ref{eqn:NS_2}) and (\ref{eqn:drho_dt_2}) using a hybrid lattice-Boltzmann method (LBM) \cite{gsell22}.
Space and time are discretized using a constant time step $\Delta t$ and grid spacing $\Delta x$. 
Equation (\ref{eqn:drho_dt_2}) is integrated using a finite volume approach with a first-order explicit Euler integration in time.
The advective term is treated using a first-order upwind scheme and the diffusive term is treated using a second-order central scheme. 
To couple the disk motion with the fluid dynamics, we use an explicit immersed-boundary method (IBM) \cite{gsell19}. 
The disk surface is discretized using Lagrangian markers on which the fluid velocity is interpolated at each time step.
A momentum correction, ensuring no-slip condition at the disk surface, is then computed along the disk surface and spread to the neighboring fluid nodes.
This momentum correction also provides the resulting hydrodynamic force $F_x$ exerted by the fluid on the particle, used in the ludion equation of motion (\ref{eqn:PFD}).
The IBM framework is also used to impose a zero-flux condition for the density at the disk surface, as follows: (i) at each time step, we use the IBM interpolation kernel to interpolate diffusive fluxes at the disk surface; (ii) we then compute the flux correction that needs to be applied in order to achieve zero normal flux and we spread this correction to neighboring fluid nodes.
Wall conditions at the edges of the computational domain are implemented by imposing $\rho = \rho_i$ (i.e. $\tilde{\rho} = 0$) at the boundaries.

\begin{figure}[b]
  \includegraphics[width=0.5\textwidth]{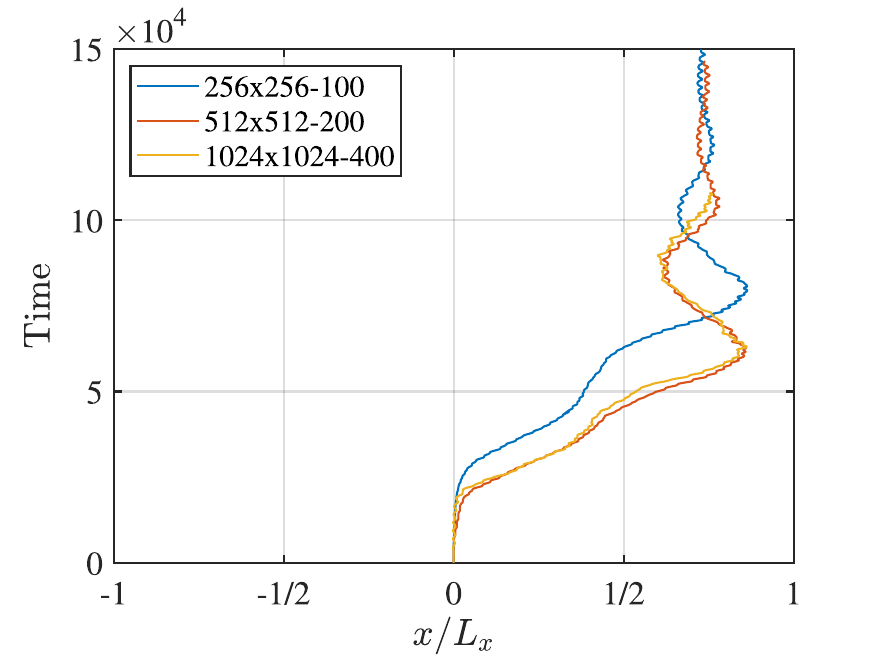}
  \caption{Ludion trajectories for $\omega / N = 0.5$ and three different mesh sizes. The grid resolution is changed together with the time step $\Delta t$ and the number of IBM markers representing the ludion geometry. Mesh 1: 256x256 grid points, 100 IBM markers, $\Delta t = T / 2400$. Mesh 2: 512x512 grid points, 200 IBM markers, $\Delta t = T / 4800$. Mesh 3: 1024x1024 grid points, 400 IBM markers, $\Delta t = T / 9600$.}
  \label{fig:mesh}
\end{figure}

Details on the numerical schemes and their performance can be found in refs. \cite{gsell22} and \cite{gsell19}.
We performed a series of specific convergence checks in order to select the mesh size and time step for the present study.
This is illustrated in \autoref{fig:mesh}, which shows the ludion trajectory in a wall-bounded domain, for $\omega/N= 0.5$ and three different mesh sizes. 
The grid resolution is changed together with the time step $\Delta t$ and the number of IBM markers representing the ludion geometry. 
Despite the highly complex dynamics, we observe no significant difference in the trajectories obtained with the 512x512 and 1024x1024 meshes. 
We therefore chose the 512x512 mesh (together with 200 IBM markers on the ludion boundary) to perform our simulations. The corresponding time step is $\Delta t = T / 4800$, with $T$ the oscillation period of the ludion.

\bibliography{ludion.bib}

\end{document}